# Constructing bibliometric networks:
# A comparison between full and fractional counting

Antonio Perianes-Rodriguez[1], Ludo Waltman[2], and Nees Jan van Eck[2]

[1] SCImago Research Group, Departamento de Biblioteconomia y Documentacion,
Universidad Carlos III, Getafe, Madrid, Spain
aperiane@bib.uc3m.es

[2] Centre for Science and Technology Studies, Leiden University, Leiden, The Netherlands
{waltmanlr, ecknjpvan}@cwts.leidenuniv.nl

The analysis of bibliometric networks, such as co-authorship, bibliographic coupling, and co-citation networks, has received a considerable amount of attention. Much less attention has been paid to the construction of these networks. We point out that different approaches can be taken to construct a bibliometric network. Normally the full counting approach is used, but we propose an alternative fractional counting approach. The basic idea of the fractional counting approach is that each action, such as co-authoring or citing a publication, should have equal weight, regardless of for instance the number of authors, citations, or references of a publication. We present two empirical analyses in which the full and fractional counting approaches yield very different results. These analyses deal with co-authorship networks of universities and bibliographic coupling networks of journals. Based on theoretical considerations and on the empirical analyses, we conclude that for many purposes the fractional counting approach is preferable over the full counting one.

## 1. Introduction

The study of bibliometric networks, such as co-authorship, bibliographic coupling, and co-citation networks, has a long history in the field of bibliometrics, with early work dating back to the 1960s and 1970s (e.g., De Solla Price, 1965; Kessler, 1963; Small, 1973). Many different methods for analyzing and visualizing bibliometric networks have been studied by bibliometricians (e.g., Börner, Chen, & Boyack, 2003; Milojević, 2014; Van Eck & Waltman, 2014; Zhao & Strotmann, 2015). However, before bibliometric networks can be analyzed and visualized, they first need to be constructed. The construction of bibliometric networks has received remarkably little attention in the literature (for important exceptions, see Batagelj & Cerinšek, 2013; Park, Yoon, & Leydesdorff, 2016). It seems that the construction of bibliometric networks is typically seen as a more or less trivial step that does not need any special consideration. In this paper, we argue that this step is far from trivial. We point out that different approaches can be taken to construct bibliometric networks. Our aim is to draw attention to the existence of different approaches for constructing bibliometric networks, to clarify the conceptual differences between these approaches, and to show that these approaches may yield very different results.

A well-known problem in the field of bibliometrics is the issue of assigning co-authored publications to individual authors. For instance, when a publication is co-authored by three researchers, how should the publication be counted for each individual researcher? In the context of the calculation of bibliometric indicators, many different approaches have been proposed to this problem (for overviews, see Gauffriau, Larsen, Maye, Roulin-Perriard, & Von Ins, 2007; Waltman, 2016, Section 7). The most popular approaches are the full counting method (also known as the whole counting method) and the fractional counting method (e.g., Aksnes, Schneider, & Gunnarsson, 2012; Waltman & Van Eck, 2015). In the case of the full counting method, a publication co-authored by three researchers is assigned to each researcher with a full weight of one. On the other hand, in the case of the fractional counting method, the publication is assigned to each researcher with a fractional weight of 1 / 3.

In this paper, we show how the distinction between full and fractional counting, which has been studied extensively in the context of the calculation of bibliometric indicators, can be translated to the context of the construction of bibliometric networks. Consider for instance the construction of a co-authorship network. Suppose researcher X has co-authored a publication with five other researchers. In the conventional approach to the construction of bibliometric networks, this yields five co-authorship links with a weight of one for researcher X. We refer to this approach as the full counting method. An alternative approach is to assign a weight of 1 / 5 to each of the five co-authorship links. In this approach, which we refer to as the fractional counting method, the total weight of the co-authorship links that a researcher obtains because of co-authoring a publication equals one. This total weight of one is distributed equally over the individual co-authorship links.

To construct bibliometric networks, researchers have traditionally used the full counting method. To the best of our knowledge, the fractional counting method has hardly been used in the literature (for the only exception that we are aware of, see Newman, 2001c), although some related ideas have been proposed (Batagelj & Cerinšek, 2013; Cerinšek & Batagelj, 2015; Park et al., 2016; Persson, 1994, 2010).[1] In this paper, we carefully define the full and fractional counting methods. Our focus is on three popular types of bibliometric networks, namely co-authorship, bibliographic coupling, and co-citation networks, but our ideas extend to other types of bibliometric networks as well. We also provide two examples of situations in which the choice between the full and fractional counting methods makes a big difference. One example is about co-authorship networks of universities. The other example deals with bibliographic coupling networks of journals. In both examples, we argue that the fractional counting method is preferable over the full counting method.

We note that the full and fractional counting methods are both available in the VOSviewer software (www.vosviewer.com; Van Eck & Waltman, 2010, 2014) for constructing and visualizing bibliometric networks. The VOSviewer software can be used to construct bibliometric networks based on data downloaded from bibliographic databases such as Web of Science and Scopus. The software requests the user to choose between the use of the full and the fractional counting method. The information provided in this paper should help VOSviewer users in choosing the most appropriate counting method for their analyses.

This paper is organized as follows. Formal definitions of the full and fractional counting methods in the context of the construction of bibliometric networks are provided in Section 2. An empirical comparison between the two counting methods is reported in Section 3. We present our conclusions in Section 4.

---

[1] Small and Sweeney (1985) also use a fractional counting approach in the context of the construction of a bibliometric network. However, they do not use fractional counting in the actual construction of the network, but instead they use fractional counting to select the publications to be included in the network.

## 2. Constructing bibliometric networks

In this section, we provide a detailed discussion of the full and fractional counting methods for constructing bibliometric networks. We first discuss in general terms the difference between full and fractional counting. We then focus specifically on co-authorship networks, followed by bibliographic coupling and co-citation networks. We focus on these three types of bibliometric networks because they seem to be the types of bibliometric networks that receive most attention in the literature. However, we emphasize that our ideas apply to other types of bibliometric networks as well. For an overview of the literature on different types of bibliometric networks, we refer to Van Eck and Waltman (2014, Subsection 2.1).

### 2.1. Full counting vs. fractional counting

In the context of the calculation of bibliometric indicators, the concepts of a publication and a co-author play a key role in the distinction between full and fractional counting. Full counting means that a co-authored publication is counted with a full weight of one for each co-author, which implies that the overall weight of a publication is equal to the number of authors of the publication. Fractional counting means that a co-authored publication is assigned fractionally to each of the co-authors, with the overall weight of the publication being equal to one. Hence, in the case of fractional counting, each publication has the same overall weight.

In the context of the construction of bibliometric networks, a similar distinction between full and fractional counting can be made. However, in order to do so, the concepts of a publication and a co-author need to be replaced by appropriate network-related concepts. We replace the concept of a publication by the concept of an action. The concept of a co-author is replaced by the concept of a link. For specific types of bibliometric networks, the concepts of an action and a link can be given a more concrete interpretation. For instance, in the case of a co-authorship network, co-authoring a publication with other researchers is an action and this action results in co-authorship links. In the case of a bibliographic coupling or co-citation network, giving a citation is an action and this action results in bibliographic coupling or co-citation links.

When full counting is used to construct a bibliometric network, each link resulting from an action has a full weight of one, which means that the overall weight of an action is equal to the number of links resulting from the action. On the other hand, when fractional counting is used, each link has a fractional weight such that the overall weight of an action equals one. For instance, in the case of fractional counting, the decision of a researcher to co-author a publication with five other researchers should have the same weight as the decision of a researcher to co-author a publication with 500 other researchers. In the first situation, five new co-authorship links are introduced. Each of these links is assigned a fractional counting weight of 1 / 5, so that the total weight equals $5 \times (1 / 5) = 1$. The second situation results in 500 new co-authorship links, each with a fractional counting weight of 1 / 500, which again yields a total weight of $500 \times (1 / 500) = 1$. In the case of full counting, each co-authorship link has a weight of one in both situations, resulting in a total weight of 5 in the first situation and 500 in the second situation. Hence, based on full counting, the decision made in the second situation has 100 times as much weight as the decision made in the first situation.

Table 1. Summary of the key differences between full and fractional counting, both in the context of the calculation of bibliometric indicators (where *N* denotes the number of co-authors of a publication) and in the context of the construction of bibliometric networks (where *N* denotes the number of links resulting from an action).

|  | Full counting | Fractional counting |
| --- | --- | --- |
| Indicators | Each *co-author* has a weight of 1. Each *publication* has a total weight of *N*. | Each *co-author* has a weight of 1 / *N*. Each *publication* has a total weight of 1. |
| Networks | Each *link* has a weight of 1. Each *action* has a total weight of *N*. | Each *link* has a weight of 1 / *N*. Each *action* has a total weight of 1. |

A completely analogous example can be given for the construction of a bibliographic coupling network, where links are created when two publications both cite the same third publication (Kessler, 1963). In the case of fractional counting, giving a citation to a publication that has already been cited by five other publications has the same weight as giving a citation to a publication that has already been cited by 500 other publications. In the first situation, five new bibliographic coupling links are introduced, each with a fractional counting weight of 1 / 5, which gives a total weight of $5 \times (1 / 5) = 1$. The second situation results in 500 new bibliographic coupling links, each with a fractional counting weight of 1 / 500, and again a total weight of $500 \times (1 / 500) = 1$ is obtained. In the case of full counting, all bibliographic coupling links have a weight of one in both situations, and therefore the total weight equals 5 in the first situation and 500 in the second situation.

The key differences between full and fractional counting are summarized in Table 1. The table also shows how full and fractional counting in the context of the construction of bibliometric networks relate to full and fractional counting in the context of the calculation of bibliometric indicators.

**2.2. Arguments in favor of fractional counting**

In the context of the construction of bibliometric networks, why would fractional counting be preferable over full counting, at least for certain purposes? In other words, why would it be reasonable to require each action to have the same weight? Let us provide an argument in the context of bibliographic coupling analysis. Suppose we have a publication and suppose we want to use bibliographic coupling analysis to identify other related publications. Bibliographic coupling analysis starts from the idea that the references cited in a publication reflect what the publication is about and, consequently, that publications citing the same references are related to each other. In the case of full counting, references that are cited not only by our focal publication but also by many other publications have a larger overall influence on the bibliographic coupling analysis than references that are cited by just a few other publications. In a certain sense, this means that in the full counting case highly cited references are seen as more representative of what a publication is about than lowly cited references. This may not be desirable.

Suppose for instance that our focal publication cites both a lowly cited research article dealing with a closely related topic and a highly cited review article that offers a broad overview of the literature, including many topics that are only weakly related to the topic of our focal publication. In this situation, the lowly cited research article is more representative of what our focal publication is about than the highly cited review article. However, in the full counting case, the reference to the highly cited review article has a much larger influence on the bibliographic coupling analysis than the reference to the lowly cited research article. One could therefore say that the reference to the highly cited review article is treated as being more representative of the topic of our focal publication than the reference to the lowly cited research article, while it actually should have been the other way around.

In the case of fractional counting, each reference cited in a publication has the same influence in a bibliographic coupling analysis, which essentially means that each reference is considered to be equally representative of what the publication is about. We believe this to be a very reasonable idea, more reasonable than the idea of highly cited references being more representative than lowly cited references. In practice, some references cited in a publication are of course more representative of what the publication is about than others. However, we see no reason to expect highly cited references to be systematically more representative than lowly cited references. Without any further information, the most reasonable idea seems to be to treat each reference cited in a publication as being equally representative, and this is what is done by fractional counting.

The above argument in favor of fractional counting applies to bibliographic coupling analysis, but similar arguments can be given for other types of analysis as well. For instance, when co-authorship analysis is used to identify strong collaborative ties between researchers, it can be argued that the most reasonable approach is to consider each publication of a researcher to be equally important in the researcher's oeuvre. This may then result in fractional counting being preferable over full counting.

**2.3. Co-authorship networks**

We now discuss in more detail the construction of co-authorship networks using full and fractional counting. We first provide a technical discussion, we then present a simple example, and finally we briefly refer to some related work in the literature.

*Constructing co-authorship networks*

Co-authorship networks can be constructed for different units of analysis, such as researchers, research institutions, and countries. In the discussion below, we use researchers as the unit of analysis (e.g., Newman, 2001a, 2001b, 2001c). However, we emphasize that the discussion also applies to other units of analysis.

We use $N$ and $M$ to denote, respectively, the number of researchers and the number of publications included in the analysis, and we use $\mathbf{A} = [a_{ik}]$ to denote an $N \times M$ authorship matrix. Element $a_{ik}$ of this matrix equals 1 if researcher $i$ is an author of publication $k$ and 0 otherwise. We further use $n_k$ to denote the number of authors of publication $k$, that is,

$$n_k = \sum_{i=1}^{N} a_{ik}. \tag{1}$$

Publications that have only one author do not provide any co-authorship links. For simplicity, we therefore assume that each publication included in the analysis has at least two authors. This means that $n_k > 1$ for each publication $k$.

We first consider the case of full counting. We use $\mathbf{U} = [u_{ij}]$ to denote the full counting co-authorship matrix. This is a symmetrical $N \times N$ matrix. Element $u_{ij}$ of this matrix equals the number of full counting co-authorship links between researchers $i$ and $j$ and is given by

$$u_{ij} = \sum_{k=1}^{M} a_{ik} a_{jk}. \tag{2}$$

In matrix notation, the co-authorship matrix $\mathbf{U}$ is given by

$$\mathbf{U} = \mathbf{A}\mathbf{A}^{\mathrm{T}}. \tag{3}$$

Hence, the co-authorship matrix $\mathbf{U}$ is obtained by post-multiplying the authorship matrix $\mathbf{A}$ by its transpose. Self-links in a co-authorship network are usually of no interest, and therefore the main diagonal elements of the co-authorship matrix $\mathbf{U}$ are set to 0.

We now consider the case of fractional counting, where we denote the fractional counting co-authorship matrix by $\mathbf{U}^* = [u^*_{ij}]$. The number of fractional counting co-authorship links between researchers $i$ and $j$, denoted by $u^*_{ij}$, is given by

$$u^*_{ij} = \sum_{k=1}^{M} \frac{a_{ik} a_{jk}}{n_k - 1}. \tag{4}$$

Equivalently, the co-authorship matrix $\mathbf{U}^*$ is obtained by

$$\mathbf{U}^* = \mathbf{A}\,\mathrm{diag}(\mathbf{A}^{\mathrm{T}}\mathbf{1} - \mathbf{1})^{-1}\,\mathbf{A}^{\mathrm{T}}, \tag{5}$$

where diag(**v**) denotes a diagonal matrix with the elements of the vector **v** on the main diagonal and where **1** denotes a column vector of length *N* with all elements equal to 1. The main diagonal elements of the co-authorship matrix $\mathbf{U}^*$ are set to 0.

*Example*

To illustrate the use of full and fractional counting for constructing co-authorship networks, we consider a simple example in which we have four researchers and three publications. Table 2 presents the authorship matrix and Figure 1 displays the corresponding authorship network.

Table 2. Authorship matrix.

|  | P1 | P2 | P3 | Total |
|---|---|---|---|---|
| R1 | 1 | 1 | 0 | 2 |
| R2 | 1 | 0 | 1 | 2 |
| R3 | 1 | 1 | 0 | 2 |
| R4 | 0 | 0 | 1 | 1 |
| Total | 3 | 2 | 2 |  |

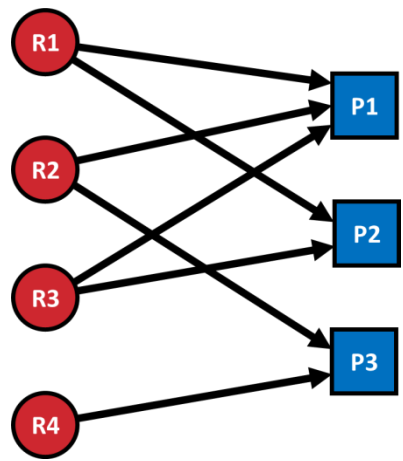

Figure 1. Authorship network.

The full and fractional counting co-authorship matrices and the corresponding co-authorship networks are presented in Table 3 and Figure 2, respectively. We note that for each researcher the total weight of the fractional counting co-authorship links is equal to the number of publications the researcher has authored. This is a general property of fractional counting co-authorship analyses.

Table 3. Full and fractional counting co-authorship matrices.

| | Full counting | | | | | | Fractional counting | | | | |
|---|---|---|---|---|---|---|---|---|---|---|---|
| | R1 | R2 | R3 | R4 | Total | | R1 | R2 | R3 | R4 | Total |
| R1 | | 1 | 2 | 0 | 3 | R1 | | 0.5 | 1.5 | 0.0 | 2.0 |
| R2 | 1 | | 1 | 1 | 3 | R2 | 0.5 | | 0.5 | 1.0 | 2.0 |
| R3 | 2 | 1 | | 0 | 3 | R3 | 1.5 | 0.5 | | 0.0 | 2.0 |
| R4 | 0 | 1 | 0 | | 1 | R4 | 0.0 | 1.0 | 0.0 | | 1.0 |
| Total | 3 | 3 | 3 | 1 | | Total | 2.0 | 2.0 | 2.0 | 1.0 | |

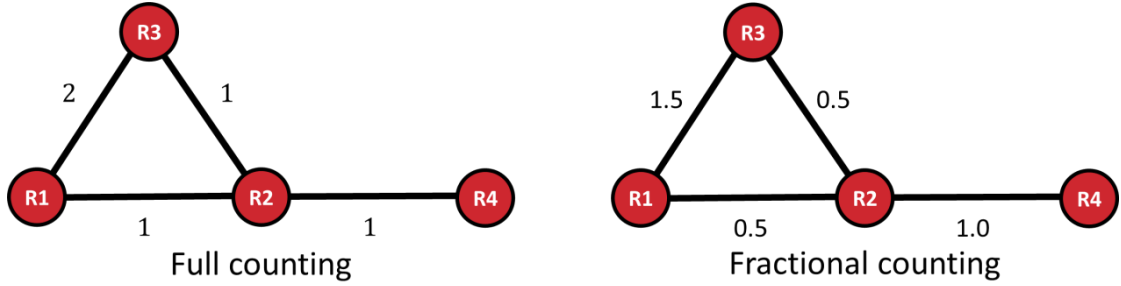

Figure 2. Full and fractional counting co-authorship networks.

To illustrate how the weights of the fractional counting co-authorship links have been obtained, we take the link between researchers 1 and 3 as an example. Researcher 1 has co-authored publication 1 with two other researchers. This yields two co-authorship links for researcher 1, and one of these links is with researcher 3. It follows from Eq. (4) that the two co-authorship links each have a weight of $1 / (3 - 1) = 0.5$. Researcher 1 has co-authored publication 2 only with researcher 3, and this results in a co-authorship link with a weight of $1 / (2 - 1) = 1$. In total, we obtain a weight of $0.5 + 1.0 = 1.5$ for the co-authorship link between researchers 1 and 3.

As explained in Subsection 2.1, in the case of fractional counting, each action should have the same weight. For instance, the decision of researcher 2 to co-author publication 1 with researchers 1 and 3 should have the same weight as researcher 2's decision to co-author publication 3 with researcher 4. The co-authorship links of researcher 2 with researchers 1 and 3 each have a weight of $1 / (3 - 1) = 0.5$, which means that the weight of researcher 2's decision to co-author publication 1 with researchers 1 and 3 equals $2 \times 0.5 = 1$. The weight of researcher 2's decision to co-author publication 3 with researcher 4 equals $1 / (2 - 1) = 1$. Hence, in the case of fractional counting, the two actions of researcher 2 indeed have the same weight.

We note that it is essential to have a denominator of $n_k - 1$ rather than $n_k$ in Eq. (4). We need to subtract 1 from $n_k$ in the denominator because we do not consider self-links in a co-authorship network. Without subtracting 1 from $n_k$, the weight of researcher 2's decision to co-author publication 1 with researchers 1 and 3 would have been $2 \times 1 / 3 = 0.67$, while the weight of researcher 2's decision to co-author publication 3 with researcher 4 would have been $1 / 2 = 0.5$. Hence, without subtracting 1 from $n_k$, the weight of the two actions of researcher 2 would not have been the same.

*Related work*

Our fractional counting method for constructing co-authorship networks is equivalent to the approach for constructing weighted co-authorship networks proposed by Newman (2001c). Our fractional counting method is also related to the approaches for constructing co-authorship networks introduced by Batagelj and Cerinšek (2013) and Park et al. (2016). In the appendix, we discuss in more detail how our fractional counting method relates to these approaches for constructing co-authorship networks.

**2.4. Bibliographic coupling networks**

In Subsection 2.3, the construction of co-authorship networks using full and fractional counting was discussed. We now turn to the construction of bibliographic coupling networks. The discussion below closely resembles the discussion in Subsection 2.3, but there are also some small differences.

*Constructing bibliographic coupling networks*

Bibliographic coupling networks can be constructed for different units of analysis, such as publications, journals, and researchers. Our focus will be on researchers as the unit of analysis (Zhao & Strotmann, 2008a), but we emphasize that the discussion below also applies to other units of analysis. In a bibliographic coupling analysis of researchers, the relatedness of researchers is determined based on the degree to which they cite the same publications. The more often two researchers cite the same publications, the stronger their relatedness.

We use $N$ and $M$ to denote, respectively, the number of researchers and the number of publications included in the analysis, and we use $\mathbf{C} = [c_{ik}]$ to denote an $N \times M$ citation matrix. Element $c_{ik}$ of this matrix equals the number of citations received by publication $k$ from researcher $i$. We further use $n_k$ to denote the total number of citations received by publication $k$ from all researchers included in the analysis, that is,

$$n_k = \sum_{i=1}^{N} c_{ik}. \tag{6}$$

Publications that have been cited fewer than two times do not provide any bibliographic coupling links. We therefore assume that each publication included in the analysis has received at least two citations, which means that $n_k > 1$ for each publication $k$.

We use $\mathbf{V} = [v_{ij}]$ to denote the $N \times N$ full counting bibliographic coupling matrix. Element $v_{ij}$ of this matrix equals the number of full counting bibliographic coupling links between researchers $i$ and $j$ and is given by

$$v_{ij} = \sum_{k=1}^{M} c_{ik} c_{jk}. \tag{7}$$

Hence, the bibliographic coupling matrix $\mathbf{V}$ is given by

$$\mathbf{V} = \mathbf{C}\mathbf{C}^\mathrm{T}. \tag{8}$$

Turning now to the fractional counting case, we use $\mathbf{V}^* = [v^*_{ij}]$ to denote the fractional counting bibliographic coupling matrix. The number of fractional counting bibliographic coupling links between researchers $i$ and $j$, denoted by $v^*_{ij}$, is given by

$$v^*_{ij} = \sum_{k=1}^{M} \frac{c_{ik} c_{jk}}{n_k - 1}. \tag{9}$$

Equivalently, the bibliographic coupling matrix $\mathbf{V}^*$ is obtained by

$$\mathbf{V}^* = \mathbf{C}\,\mathrm{diag}(\mathbf{C}^\mathrm{T}\mathbf{1} - \mathbf{1})^{-1}\,\mathbf{C}^\mathrm{T}. \tag{10}$$

Self-links in a bibliographic coupling network are usually of no interest, and therefore the main diagonal elements of the bibliographic coupling matrices **V** and **V**$^*$ are set to 0.

*Example*

We consider an example with five researchers and four publications. The citation matrix and the corresponding citation network are presented in Table 4 and Figure 3, respectively. We note that a researcher can give multiple citations to the same publication. For instance, researcher 1 has cited publication 1 three times. This means that researcher 1 has authored three publications in which publication 1 is cited.

Table 4. Citation matrix.

|       | P1 | P2 | P3 | P4 | Total |
|-------|----|----|----|----|-------|
| R1    | 3  | 1  | 2  | 0  | 6     |
| R2    | 2  | 0  | 1  | 0  | 3     |
| R3    | 1  | 2  | 0  | 0  | 3     |
| R4    | 0  | 0  | 0  | 1  | 1     |
| R5    | 0  | 1  | 0  | 1  | 2     |
| Total | 6  | 4  | 3  | 2  |       |

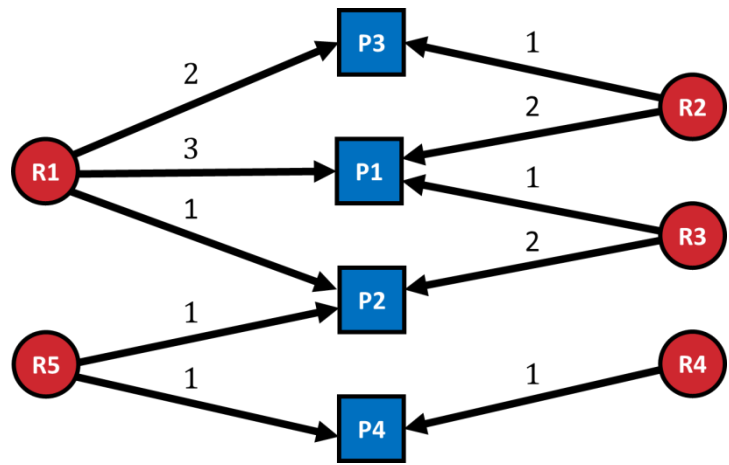

Figure 3. Citation network.

The full and fractional counting bibliographic coupling matrices and the corresponding bibliographic coupling networks can be found in Table 5 and Figure 4, respectively.

Table 5. Full and fractional counting bibliographic coupling matrices.

| | Full counting | | | | | | | Fractional counting | | | | | |
|---|---|---|---|---|---|---|---|---|---|---|---|---|---|
| | R1 | R2 | R3 | R4 | R5 | Total | | R1 | R2 | R3 | R4 | R5 | Total |
| R1 | | 8 | 5 | 0 | 1 | 14 | R1 | | 2.20 | 1.27 | 0.00 | 0.33 | 3.80 |
| R2 | 8 | | 2 | 0 | 0 | 10 | R2 | 2.20 | | 0.40 | 0.00 | 0.00 | 2.60 |
| R3 | 5 | 2 | | 0 | 2 | 9 | R3 | 1.27 | 0.40 | | 0.00 | 0.67 | 2.33 |
| R4 | 0 | 0 | 0 | | 1 | 1 | R4 | 0.00 | 0.00 | 0.00 | | 1.00 | 1.00 |
| R5 | 1 | 0 | 2 | 1 | | 4 | R5 | 0.33 | 0.00 | 0.67 | 1.00 | | 2.00 |
| Total | 14 | 10 | 9 | 1 | 4 | | Total | 3.80 | 2.60 | 2.33 | 1.00 | 2.00 | |

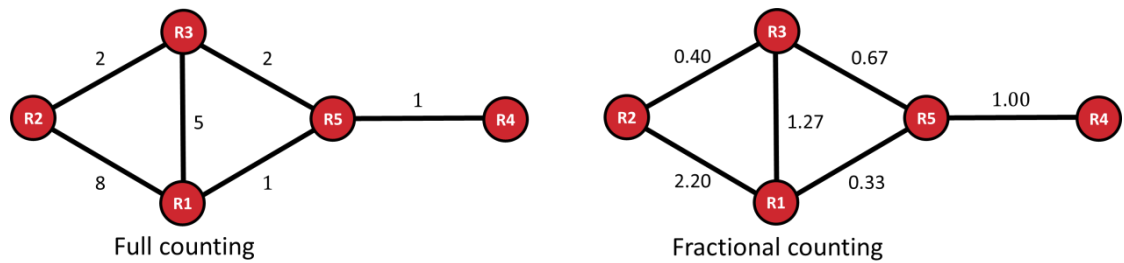

Figure 4. Full and fractional counting bibliographic coupling networks.

This example can be used to illustrate how fractional counting implements the idea that each action should have the same weight. Researcher 5 cites publication 4, which results in a bibliographic coupling link with researcher 4 with a weight of $1 / (2 - 1) = 1$. Likewise, researcher 5 cites publication 2, resulting in bibliographic coupling links with researchers 1 and 3 that have weights of, respectively, $1 / (4 - 1) = 0.33$ and $2 / (4 - 1) = 0.67$, which corresponds with a total weight of $0.33 + 0.67 = 1$. This shows that the two actions of researcher 5 both have the same weight of one.

Let us now consider researcher 3. This researcher cites publication 1, which results in bibliographic coupling links with researchers 1 and 2 that have weights of, respectively, $3 / (6 - 1) = 0.6$ and $2 / (6 - 1) = 0.4$, yielding a total weight of $0.6 + 0.4 = 1$. Researcher 3 also gives two citations to publication 2. These citations require a more detailed discussion. In total, publication 2 is cited four times. Each citation of publication 2 therefore corresponds with three bibliographic coupling links, each with a weight of $1 / 3 = 0.33$, which gives a total weight of one. However, because researcher 3 gives two citations to publication 2, one of the bibliographic coupling links that we have is a link between the two citing publications of researcher 3. Since we are not interested in researcher self-links, this link is ignored. As a consequence, for each of researcher 3's citations to publication 2, the total weight of the corresponding bibliographic coupling links is less than one. More specifically, each citation corresponds with a bibliographic coupling link with researcher 1 and a bibliographic coupling link with researcher 5, and these links each have a weight of 1 / 3 = 0.33, yielding a total weight of 2 × 0.33 = 0.67. Hence, if researcher self-links had been taken into consideration, a total weight of one would have been obtained, but by ignoring researcher self-links we obtain a total weight below one.[2] This also explains why for some researchers (i.e., researchers 1, 2, and 3) the total weight of their fractional counting bibliographic coupling links is less than the number of citations they have made.

*Related work*

We are not aware of earlier work discussing approaches for constructing bibliographic coupling networks similar to our fractional counting method. The most closely related work seems to be the approach proposed by Batagelj and Cerinšek (2013) for constructing 'normalized' bibliographic coupling networks. Like our fractional counting method, the approach of Batagelj and Cerinšek (2013) is based on the idea of fractionalization. However, there is a fundamental difference. While we fractionalize based on the number of citations received by a cited publication from other publications, Batagelj and Cerinšek (2013) fractionalize based on the number of citations given by a citing publication to other publications.[3]

**2.5. Co-citation networks**

After discussing the construction of co-authorship and bibliographic coupling networks using full and fractional counting, we now consider the construction of co-citation networks. Since the construction of co-citation networks is very similar to the construction of co-authorship and bibliographic coupling networks, only a brief discussion will be provided.

---

[2] If this is considered undesirable, it can be fixed by adapting the denominator in Eq. (9). If in the denominator we subtract $c_{ik}$ rather than 1 from $n_k$, we always obtain a total weight of one. However, the bibliographic coupling matrix $\mathbf{V}^*$ may no longer be symmetrical when this approach is taken.

[3] A somewhat similar approach is taken by Sen and Gan (1983) and Glänzel and Czerwon (1996). These authors also perform a normalization based on the number of citations given by a citing publication to other publications.

*Constructing co-citation networks*

Our focus will be on researchers as the unit of analysis (McCain, 1990; White & Griffith, 1981), but we emphasize that the discussion below also applies to other units of analysis, such as publications and journals. In a co-citation analysis of researchers, the relatedness of researchers is determined based on the degree to which they are cited in the same publications. The more often two researchers are cited in the same publications, the stronger their relatedness.

Like in Subsection 2.4, we use $N$ and $M$ to denote, respectively, the number of researchers and the number of publications included in the analysis, and we use $\mathbf{C} = [c_{ik}]$ to denote an $N \times M$ citation matrix. Importantly, however, the citation matrix is defined in a different way than in Subsection 2.4. Element $c_{ik}$ of the matrix equals the number of citations given by publication $k$ to researcher $i$ (rather than the number of citations received by publication $k$ from researcher $i$). We further use $n_k$ to denote the total number of citations given by publication $k$ to all researchers included in the analysis, that is,

$$n_k = \sum_{i=1}^{N} c_{ik}. \tag{11}$$

We assume that $n_k > 1$ for each publication $k$.

Apart from the difference in the definition of the citation matrix $\mathbf{C}$, co-citation analysis is mathematically identical to bibliographic coupling analysis. We use $\mathbf{W} = [w_{ij}]$ to denote the $N \times N$ full counting co-citation matrix. Element $w_{ij}$ of this matrix equals the number of full counting co-citation links between researchers $i$ and $j$ and is given by

$$w_{ij} = \sum_{k=1}^{M} c_{ik} c_{jk}. \tag{12}$$

The co-citation matrix $\mathbf{W}$ is given by

$$\mathbf{W} = \mathbf{CC}^{\mathrm{T}}. \tag{13}$$

In the fractional counting case, we use $\mathbf{W}^* = [w^*_{ij}]$ to denote the fractional counting co-citation matrix. The number of fractional counting co-citation links between researchers $i$ and $j$, denoted by $w^*_{ij}$, is given by

$$w^*_{ij} = \sum_{k=1}^{M} \frac{c_{ik} c_{jk}}{n_k - 1}. \qquad (14)$$

The co-citation matrix $\mathbf{W}^*$ is obtained by

$$\mathbf{W}^* = \mathbf{C}\, \text{diag}(\mathbf{C}^T \mathbf{1} - \mathbf{1})^{-1}\, \mathbf{C}^T. \qquad (15)$$

Self-links in a co-citation network are usually of no interest, and therefore the main diagonal elements of the co-citation matrices $\mathbf{W}$ and $\mathbf{W}^*$ are set to 0.

*Related work*

Our fractional counting method for constructing co-citation networks is somewhat similar to a method for constructing co-citation networks discussed by Persson (1994). The latter method is used to construct 'normalized' co-citation networks. One element in the normalization is a fractionalization similar to the one proposed in Eq. (14). The difference is that a denominator of $n_k$ is used instead of the denominator of $n_k - 1$ used in Eq. (14). This is analogous to the difference between our fractional counting method for constructing co-authorship networks and one of the approaches for constructing co-authorship networks discussed by Batagelj and Cerinšek (2013) (see the appendix for more details on this difference).

We further note that there has been some discussion in the literature on how to handle publications with multiple authors when constructing co-citation networks of researchers. These discussions are about the distinction between taking into account all authors of a publication or only the first or the last one (Persson, 2001; Zhao, 2006; Zhao & Strotmann, 2008b, 2011) and about the distinction between co-citation links and co-authorship links (Rousseau & Zuccala, 2004). We do not discuss these issues in more detail in this paper.

## 3. Empirical analysis

We now present an empirical comparison of the full and fractional counting methods for constructing bibliometric networks. We will compare the results obtained using the two counting methods, but in addition we will also show why the two counting methods yield different results. Two analyses are presented. The first analysis focuses on co-authorship networks of universities. The second analysis is about bibliographic coupling networks of journals. We have selected these two analyses because full and fractional counting yield very different results in these analyses. The analyses therefore offer important insights into the differences between the two counting methods.

### 3.1. Co-authorship networks of universities

We collected all 1.28 million publications indexed in the Web of Science database that were published in 2014 and that are authored by one or more of the 750 universities included in the 2015 edition of the CWTS Leiden Ranking (www.leidenranking.com; Waltman et al., 2012). Based on these publications, we constructed a full counting and a fractional counting co-authorship network of the 750 universities. Other institutions that have co-authored with the 750 universities were ignored in the analysis. The co-authorship networks were constructed following the calculations discussed in Subsection 2.3. The VOSviewer software (Van Eck & Waltman, 2010, 2014) was used to create visualizations of the full and fractional counting co-authorship networks.

Figures 5 and 6 present visualizations of the university co-authorship networks constructed using full and fractional counting, respectively. Each circle represents a university. To prevent the names of universities from overlapping each other, names are shown only for a subset of the universities. The size of a circle reflects the number of publications of the corresponding university. The distance between two circles approximately indicates the strength of the co-authorship link between the corresponding universities. In general, the closer two circles are located to each other, the stronger the co-authorship link between the universities. Colors represents clusters of universities with strong co-authorship links. Lines are used to indicate the 1,500 strongest co-authorship links between universities.[4]

It is evident that there are large differences between the visualizations presented in Figures 5 and 6. In Figure 5, it is hard to identify a clear pattern in the visualization. Almost all universities are located together in one big group, with the exception of universities from a number of Asian countries located in the bottom area of the visualization. No clear grouping of universities by country is visible, neither in the positioning of the universities in the visualization nor in the clustering of the universities. For instance, while many US universities are located in the left area of the visualization, where they belong to the cyan, yellow, and green clusters, US universities can also be found in the bottom-right area of the visualization, where they mostly belong to the purple cluster.

In Figure 6, on the other hand, the visualization shows a very clear pattern, both in the positioning and in the clustering of the universities. A number of distinct groups of universities are visible, and to a large extent universities turn out to be grouped by country. US universities are located in the bottom area of the visualization. In the left area, groups of Chinese, Taiwanese, Japanese, and South Korean universities can be found. In the center of the visualization, we observe an Australian and a Canadian group of universities. European universities and universities from South American countries are located in the right area of the visualization, where again a reasonably strong separation by country can be observed.

The visualizations presented in Figures 5 and 6 are based on the same underlying data, but nevertheless they give a very different impression of worldwide scientific collaboration. The visualization in Figure 6, based on fractional counting, suggests that scientific collaboration takes place mostly within national borders. On the other hand, the visualization in Figure 5, based on full counting, gives the impression that national borders play only a minor role in determining scientific collaboration. How can these large differences between the two visualizations be explained?

---

[4] To produce the visualizations using the VOSviewer software, the layout attraction and layout repulsion parameters were set to 1 and 0, respectively. The clustering resolution and minimum cluster size parameters were set to 1.25 and 5, respectively.

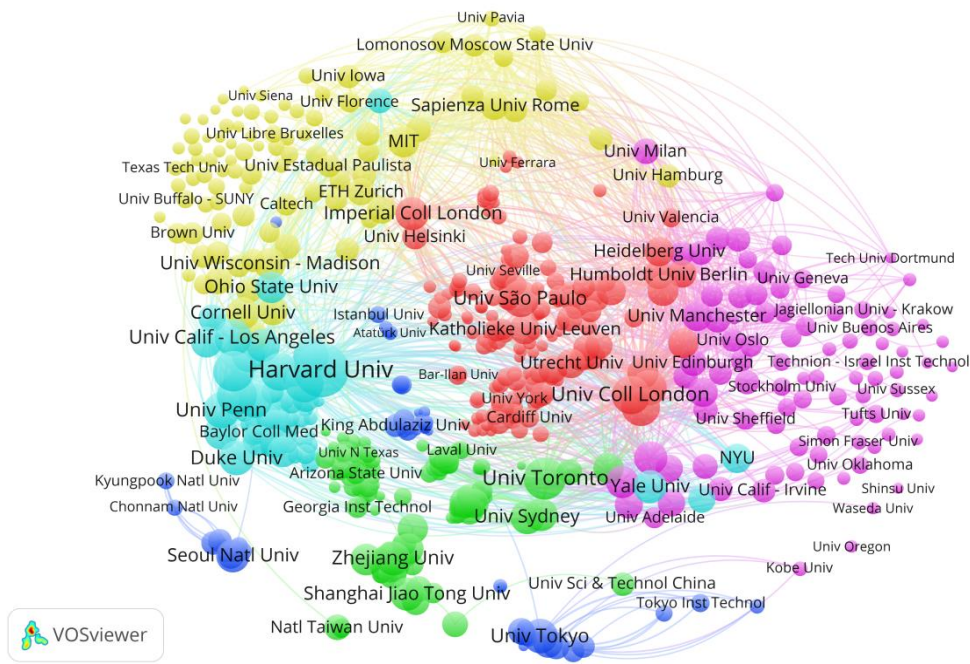

Figure 5. Visualization of the university co-authorship network constructed using full counting. An interactive visualization is available at http://goo.gl/teyI8A.

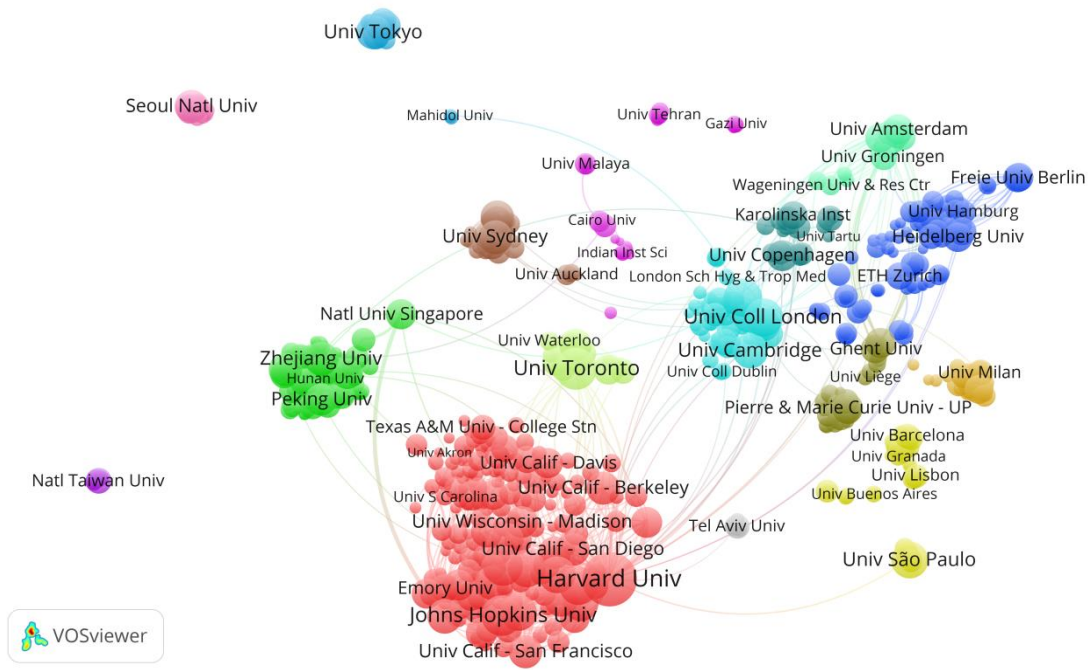

Figure 6. Visualization of the university co-authorship network constructed using fractional counting. An interactive visualization is available at http://goo.gl/wOyCEJ.

It turns out that the differences can be explained largely by the fact that in the case of full counting a small number of publications that have been co-authored by a large number of universities have a very strong effect on the co-authorship network. To demonstrate this, we constructed a full counting co-authorship network in the same way as above, except that in the construction of the network we did not take into account publications co-authored by more than 20 universities. There are 702 publications that have been co-authored by more than 20 universities (i.e., 0.05% of the total number of 1.28 million publications), and these publications were not used in the construction of the co-authorship network. A visualization of the co-authorship network that was obtained in this way is presented in Figure 7.

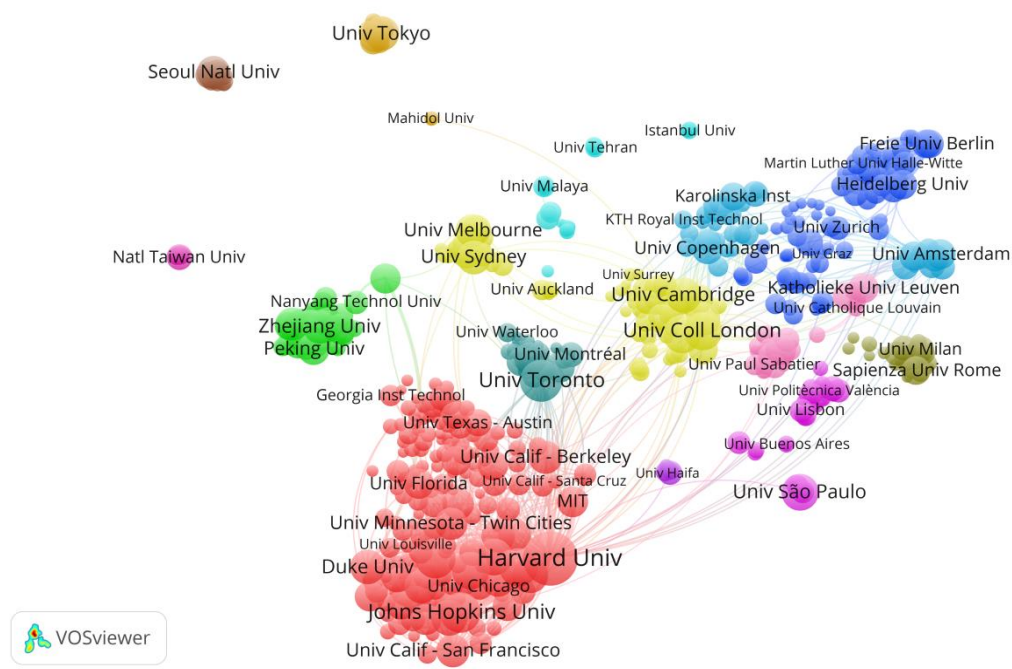

Figure 7. Visualization of the university co-authorship network constructed using full counting by including only publications co-authored by at most 20 universities. An interactive visualization is available at http://goo.gl/dgB2lT.

Importantly, the visualization in Figure 7 based on full counting is very different from the full counting visualization in Figure 5, and in fact it is quite similar to the fractional counting visualization in Figure 6. Like in the visualization in Figure 6, distinct groups of universities can be easily distinguished, and these groups largely coincide with the countries in which universities are located. Hence, it can be concluded that to a large extent the differences between full and fractional counting co-authorship networks of universities are caused by a small number of publications that have been co-authored by a large number of universities.

Table 6 provides some statistics that indicate the effect of a small number of publications with many co-authors on university co-authorship networks constructed using full counting. When in our analysis we take into account all publications regardless of their number of co-authors, we have 1.28 million publications, which yield 2.90 million co-authorship links.[5] The statistics reported in Table 6 show what happens when publications for which the number of co-authoring universities exceeds a certain threshold are not considered in the construction of a co-authorship network. In the case of the construction of the co-authorship network visualized in Figure 7, publications with more than 20 co-authoring universities were not considered. This causes a decrease of 0.05% in the number of publications. However, as can be seen in Table 6, this negligible decrease in the number of publications is responsible for a decrease of 62% in the number of co-authorship links. Even more extreme results are obtained when we take into account all publications except for those with more than 100 co-authoring universities. In that case, we lose just 0.01% of all publications, but this leads to a reduction in the number of co-authorship links by almost 50%. Based on these statistics, it is clear that in the case of full counting a very small number of publications may have a huge effect on a co-authorship network.

Table 6. Number of publications considered in the construction of a co-authorship network and number of co-authorship links included in the network when publications for which the number of co-authoring universities exceeds a certain threshold are not taken into account.

| Threshold on no. of co-authoring universities | No. of publications | % of publications | No. of co-authorship links | % of co-authorship links |
| --- | --- | --- | --- | --- |
| 5 | 1,266,634 | 99.05% | 722,935 | 25% |
| 10 | 1,276,318 | 99.80% | 939,667 | 32% |
| 20 | 1,278,123 | 99.95% | 1,102,564 | 38% |
| 50 | 1,278,585 | 99.98% | 1,372,300 | 47% |
| 100 | 1,278,667 | 99.99% | 1,532,105 | 53% |
| No threshold | 1,278,825 | 100.00% | 2,898,820 | 100% |

---

[5] If two universities have co-authored 100 publications, this can be counted either as 100 unweighted co-authorship links or as one weighted co-authorship link, where the weight equals 100. We here count co-authorship links using the former approach.

Figure 8 offers more detailed insight into the effect of publications co-authored by a large number of universities. We again explore the situation where publications for which the number of co-authoring universities exceeds a certain threshold are not considered in the construction of a co-authorship network. The figure shows how the percentage of the publications that are taken into account in the construction of a co-authorship network increases as we increase the threshold. Moreover, the figure also shows the effect of increasing the threshold on the percentage of all co-authorship links that are included in the network.

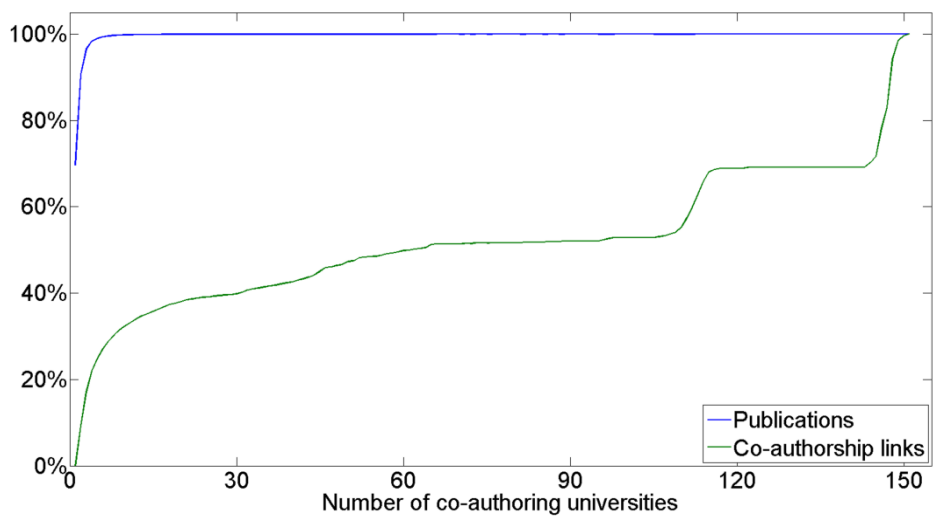

Figure 8. Percentage of publications considered in the construction of a co-authorship network and percentage of co-authorship links included in the network when publications for which the number of co-authoring universities exceeds a certain threshold are not taken into account.

Figure 8 shows that most co-authorship links are due to publications that either have a limited number of co-authoring universities or a very large number of co-authoring universities. Publications co-authored by at most ten universities are responsible for somewhat more than 30% of all co-authorship links. Individually, each of these publications contributes only a very small number of co-authorship links. However, because there are so many publications co-authored by at most ten universities (i.e., 99.8% of all publications), these publications are still responsible for almost one-third of all co-authorship links. We note that most publications (i.e., almost 70% of all publications) have been authored by just one university. These publications do not result in any co-authorship links at all.

Publications co-authored by more than 100 universities are responsible for almost 50% of all co-authorship links. There are just 158 publications that have been co-authored by more than 100 universities, but each of these hyperauthorship publications (Cronin, 2001) is responsible for a very large number of co-authorship links. For instance, the publication co-authored by most universities is a publication that has 151 co-authoring universities[6], and this single publication results in $151 \times 150 / 2 = 11,325$ co-authorship links, which is 0.4% of all co-authorship links. The 158 publications co-authored by more than 100 universities have all appeared in the field of physics, and they all or almost all seem to result from research related to the Large Hadron Collider at CERN.

We have now seen how in the case of full counting a very small number of publications with many co-authors may have a huge effect on a co-authorship network. In the case of fractional counting, the effect of publications with many co-authors is much more limited. Fractional counting is based on the idea that each action should have the same weight. Hence, each decision of a university to co-author a publication has the same weight of one, regardless of the total number of universities by which a publication is co-authored. This means that the total weight of the co-authorship links related to a publication is equal to the number of co-authoring universities. In other words, in the fractional counting case, the effect of a publication on a co-authorship network increases linearly with the number of co-authors. In the full counting case, on the other hand, the effect of a publication increases quadratically with the number of co-authors. We have for instance seen that in the full counting case 0.05% of all publications are responsible for 62% of all co-authorship links. In the fractional counting case, the same publications turn out to be responsible for just 4.0% of all co-authorship links.

**3.2. Bibliographic coupling networks of journals**

We now turn to the analysis of bibliographic coupling networks of journals. Our aim is to use bibliographic coupling to identify the journals that are most strongly related to one specific focal journal. We use *Scientometrics* as the focal journal, since this is a journal that we expect many readers of this paper to be familiar with.

---

[6] This is the following publication: Aad et al. (2014). Search for long-lived neutral particles decaying into lepton jets in proton-proton collisions at $\sqrt{s} = 8$ Tev with the ATLAS detector. *Journal of High Energy Physics*, 11, 88.

We again performed our analysis using the Web of Science database. Following the calculations discussed in Subsection 2.4, two bibliographic coupling networks of journals were constructed, one based on full counting and one based on fractional counting. The networks were constructed based on citing publications in the period 2010–2014. In *Scientometrics*, 1,350 publications appeared in this period. These 1,350 citing publications refer to 12,799 publications indexed in the Web of Science database, resulting in bibliographic coupling links of *Scientometrics* with 11,526 other journals.

Table 7. The 20 journals most strongly related to *Scientometrics* in the full counting bibliographic coupling network.

| Rank | | Journal | No. of bib. coupling links | |
|---|---|---|---|---|
| Full | Frac. | | Full | Frac. |
| 1 | 1 | Journal of Informetrics | 94,561 | 1,674.8 |
| 2 | 4 | PLOS ONE | 76,369 | 518.0 |
| 3 | 2 | J. of the Am. Soc. for Information Science and Technology | 61,478 | 1,331.8 |
| 4 | 30 | Physical Review E | 43,132 | 69.6 |
| 5 | 21 | Physica A | 42,938 | 104.5 |
| 6 | 3 | Research Policy | 42,434 | 568.7 |
| 7 | 1,674 | Acta Crystallographica Section E | 22,720 | 1.7 |
| 8 | 34 | Scientific Reports | 17,649 | 62.1 |
| 9 | 6 | Technological Forecasting and Social Change | 15,228 | 336.9 |
| 10 | 28 | Strategic Management Journal | 14,025 | 70.1 |
| 11 | 7 | J. of the Ass. for Information Science and Technology | 13,901 | 308.6 |
| 12 | 5 | Research Evaluation | 13,107 | 348.8 |
| 13 | 12 | Technovation | 12,831 | 162.8 |
| 14 | 39 | Organization Science | 12,829 | 57.8 |
| 15 | 9 | Journal of Technology Transfer | 12,391 | 198.7 |
| 16 | 99 | Europhysics Letters | 12,108 | 24.6 |
| 17 | 14 | Expert Systems with Applications | 10,597 | 158.1 |
| 18 | 126 | European Physical Journal B | 10,532 | 20.4 |
| 19 | 11 | Technology Analysis & Strategic Management | 10,452 | 163.6 |
| 20 | 758 | Physical Review B | 10,373 | 4.2 |

Table 7 lists the 20 journals that are most strongly related to *Scientometrics* in the full counting bibliographic coupling network. For each journal, both the number of full counting and the number of fractional counting bibliographic coupling links with *Scientometrics* is reported. Table 8 is similar to Table 7, but it shows the 20 journals that are most strongly related to *Scientometrics* in the fractional counting rather than the full counting bibliographic coupling network.

Table 8. The 20 journals most strongly related to *Scientometrics* in the fractional counting bibliographic coupling network.

| Rank | | Journal | No. of bib. coupling links | |
|---|---|---|---|---|
| Full | Frac. | | Full | Frac. |
| 1 | 1 | Journal of Informetrics | 94,561 | 1,674.8 |
| 3 | 2 | J. of the Am. Soc. for Information Science and Technology | 61,478 | 1,331.8 |
| 6 | 3 | Research Policy | 42,434 | 568.7 |
| 2 | 4 | PLOS ONE | 76,369 | 518.0 |
| 12 | 5 | Research Evaluation | 13,107 | 348.8 |
| 9 | 6 | Technological Forecasting and Social Change | 15,228 | 336.9 |
| 11 | 7 | J. of the Ass. for Information Science and Technology | 13,901 | 308.6 |
| 33 | 8 | Revista Espanola de Documentacion Cientifica | 7,848 | 204.7 |
| 15 | 9 | Journal of Technology Transfer | 12,391 | 198.7 |
| 38 | 10 | Malaysian Journal of Library & Information Science | 7,119 | 174.9 |
| 19 | 11 | Technology Analysis & Strategic Management | 10,452 | 163.6 |
| 13 | 12 | Technovation | 12,831 | 162.8 |
| 35 | 13 | Online Information Review | 7,547 | 159.7 |
| 17 | 14 | Expert Systems with Applications | 10,597 | 158.1 |
| 40 | 15 | Journal of Information Science | 6,679 | 144.6 |
| 37 | 16 | Current Science | 7,255 | 137.9 |
| 41 | 17 | Science and Public Policy | 6,560 | 127.8 |
| 32 | 18 | Information Processing & Management | 7,876 | 123.4 |
| 75 | 19 | Higher Education | 4,369 | 121.9 |
| 81 | 20 | Journal of Documentation | 3,970 | 115.5 |

As can be seen in Table 8, journals that are highly ranked based on fractional counting also tend to be quite highly ranked based on full counting. Importantly, however, Table 7 shows that this does not apply in the reverse situation. Some journals are highly ranked based on full counting, while they are ranked much lower based on fractional counting. The most extreme case is *Acta Crystallographica Section E* (*ACSE*). Based on full counting, this journal is ranked as the seventh most strongly related journal to *Scientometrics*. On the other hand, based on fractional counting, it is at rank 1,674 in terms of its relatedness with *Scientometrics*. Another extreme case is *Physical Review B*. This journal is at rank 20 based on full counting, while it is at rank 758 based on fractional counting.

Intuitively, based on our knowledge of *Scientometrics*, the full counting results, indicating a strong relatedness of *Scientometrics* with *ACSE* and *Physical Review B*, seem questionable to us. Let us therefore analyze why full counting yields these results and why fractional counting gives results that are so much different. We present an analysis for *ACSE*, because this is the journal for which the differences between the full and fractional counting results are most extreme. However, analyses for other journals for which there is large difference provide similar insights.

There turn out to be just ten publications that in the period 2010–2014 were cited both by *ACSE* and by *Scientometrics*. Hence, all bibliographic coupling links between *ACSE* and *Scientometrics* are due to these ten cited publications. The ten publications are listed in Table 9. For each publication, the table reports the number of times the publication was cited by *ACSE* and by *Scientometrics* in the period 2010–2014. The table also presents for each publication the resulting number of full counting bibliographic coupling links, which is obtained by multiplying the number of citations received from *ACSE* by the number of citations received from *Scientometrics*. As can also be seen in Table 7, in total there are 22,720 bibliographic coupling links between *ACSE* and *Scientometrics*.

The first two publications listed in Table 9 turn out to be responsible for 99.9% of all full counting bibliographic coupling links between *ACSE* and *Scientometrics*. Interestingly, these two publications have each been cited just two times by *Scientometrics*. However, they have both received a very large number of citations from *ACSE*, and this explains why in the case of full counting these two publications result in a very large number of bibliographic coupling links between *ACSE* and *Scientometrics*. Hence, the fact that based on full counting *ACSE* is the seventh most strongly related journal to *Scientometrics* is due to just four citations given by *Scientometrics*. These citations happen to refer to publications that have been cited a lot by *ACSE*, and therefore they result in a very large number of bibliographic coupling links, making *ACSE* the seventh most strongly related journal to *Scientometrics*.

Table 9. The ten cited publications responsible for all bibliographic coupling links between *Acta Crystallographica Section E* and *Scientometrics*.

| Publication | No. of citations | | No. of bib. coupling links |
|---|---|---|---|
| | ACSE | Scientom. | |
| Sheldrick, G.M. (2008). A short history of SHELX. *Acta Crystallographica Section A*, *64*, 112–122. | 8,659 | 2 | 17,318 |
| Spek A.L. (2009). Structure validation in chemical crystallography. *Acta Crystallographica Section D*, *65*, 148–155. | 2,687 | 2 | 5,374 |
| Desiraju, G.R. (2002). Hydrogen bridges in crystal engineering: Interactions without borders. *Accounts of Chemical Research*, *35*(7), 565–573. | 8 | 1 | 8 |
| Desiraju, G.R. (1995). Supramolecular synthons in crystal engineering – A new organic synthesis. *Angewandte Chemie*, *34*(21), 2311–2327. | 6 | 1 | 6 |
| Desiraju, G.R. (1996). The C-H...O hydrogen bond: Structural implications and supramolecular design. *Accounts of Chemical Research*, *29*(9), 441–449. | 5 | 1 | 5 |
| Becke A.D. (1993). Density-functional thermochemistry. III. The role of exact exchange. *Journal of Chemical Physics*, *98*, 5648. | 4 | 1 | 4 |
| Kroto, H.W. et al. (1985). $C_{60}$: Buckminsterfullerene. *Nature*, *318*, 162–163. | 2 | 1 | 2 |
| De Clercq, E. (2009). The history of antiretrovirals: Key discoveries over the past 25 years. *Reviews in Medical Virology*, *19*(5), 287–299. | 1 | 1 | 1 |
| Desiraju, G.R. (1991). The C-H...O hydrogen bond in crystals: What is it? *Accounts of Chemical Research*, *24*(10), 290–296. | 1 | 1 | 1 |
| Hoeben, F.J.M. et al. (2005). About supramolecular assemblies of pi-conjugated systems. *Chemical Reviews*, *105*(4), 1491–1546. | 1 | 1 | 1 |

In the case of fractional counting, each action has equal weight. This means that a citation to a highly cited publication and a citation to a lowly cited publication each result in bibliographic coupling links with the same total weight. The first publication listed in Table 9 in total was cited 30,798 times in the period 2010–2014 (not only by *ACSE* and by *Scientometrics* but also by other journals). For each citing publication, this yields 30,798 − 1 = 30,797 bibliographic coupling links with other citing publications. In the fractional counting case, each of these bibliographic coupling links has a weight of 1 / 30,797 = 0.000032. This means that the total weight of the bibliographic coupling links between *ACSE* and *Scientometrics* resulting from the first publication in Table 9 equals 2 × 8,659 × 0.000032 = 0.56. Likewise, the second publication in Table 9 in total was cited 4,930 times in the period 2010–2014, which results in bibliographic coupling links between *ACSE* and *Scientometrics* with a total weight of 2 × 2,687 × (1 / (4,930 − 1)) = 1.09. In the fractional counting case, the overall weight of all bibliographic coupling links between *ACSE* and *Scientometrics* turns out to be rather low, and therefore *ACSE* ends up at rank 1,674 in terms of its relatedness with *Scientometrics*.

The above analysis shows how full and fractional counting may lead to completely different results in determining the relatedness of journals based on bibliographic coupling. The analysis focuses on the relatedness of *Scientometrics* with *ACSE*, but similar insights can be obtained from an analysis of the relatedness of *Scientometrics* with for instance *Physical Review B*. We consider the fractional counting results to be more useful than the full counting ones. The full counting results match less well with our intuitive idea of the relatedness of *Scientometrics* with other journals, and the strong relatedness of *Scientometrics* with *ACSE*, which is based on just four citations given by *Scientometrics*, does not make much sense to us.

## 4. Discussion and conclusion

There is an extensive literature on the study of bibliometric networks, for instance to describe the structure and evolution of scientific collaboration, to detect research fronts, and to identify specialties within a discipline. However, with a few exceptions (Batagelj & Cerinšek, 2013; Park et al., 2016), the literature pays hardly any attention to the construction of bibliometric networks. Constructing a bibliometric network is seen as a more or less trivial step, and there seems to be an implicit idea that the only way to construct a bibliometric network is to use the full counting method.

In this paper, we have argued that there are different approaches that can be taken in the construction of a bibliometric network, and in particular we have emphasized the distinction between approaches based on full and fractional counting. We have pointed out that the fractional counting method has the attractive property that each action, such as co-authoring or citing a publication, has equal weight. In the case of the full counting method, some actions may have much more weight than others, which we believe may not be desirable.

**4.1. How much difference does the choice of a counting method make?**

When constructing a bibliometric network, how much difference does the choice between full and fractional counting make in practice? Our experience is that in many cases the differences are relatively limited and may not have a fundamental effect on the conclusions drawn from the analysis of a bibliometric network. Especially in analyses based on small data sets, it seems unlikely that results obtained using full and fractional counting will be very different. However, researchers increasingly perform analyses based on large data sets, and it then seems more likely that there will be substantial differences between results obtained using the two counting methods. In this paper, we have provided two examples of analyses in which the choice between full and fractional counting does indeed make a big difference.

In the first example, we have demonstrated how the choice between full and fractional counting has a strong effect on co-authorship networks of universities. A visualization of a full counting co-authorship network of universities suggests that national borders play only a minor role in determining scientific collaboration, while a visualization of the corresponding fractional counting network gives the impression that scientific collaboration is strongly organized within national borders. As we have shown, the difference is due to the fact that in the full counting case a small number of publications that have been co-authored by a large number of universities have a dominant effect on the co-authorship network. In the fractional counting case, the effect of these publications is strongly reduced.

In the second example, we have demonstrated the effect of the choice between full and fractional counting on bibliographic coupling networks of journals. Although this example focuses on a different type of bibliometric network than the first example, the underlying mechanism that explains the differences between the results obtained using the two counting methods is similar. In the full counting case, bibliographic coupling indicates that *Scientometrics* is strongly related to *Acta Crystallographica Section E* and *Physical Review B*. In the fractional counting case, the relatedness of *Scientometrics* with these journals is much weaker. It turns out that in the full counting case a small number of citations to a few highly cited publications have a very strong influence on the outcomes of the bibliographic coupling analysis, while in the fractional counting case these citations have much less influence.

**4.2. Which counting method should be preferred?**

We do not want to argue that one counting method should always be preferred over another. Different counting methods provide different perspectives on the underlying data, and it ultimately depends on the purpose of the analysis which perspective is more useful. However, based on the theoretical considerations and the empirical analyses presented in this paper, our conclusion is that in many situations, for instance in the two examples discussed above, fractional counting offers a more useful perspective than full counting. We believe that results obtained using full counting may relatively easily lead to misunderstandings or misinterpretations. This can be avoided by using fractional counting.

On the other hand, we also want to draw attention to a disadvantage of fractional counting relative to full counting. Fractional counting is more difficult to explain than full counting. The use of fractional counting results in bibliometric networks in which links have non-integer weights. In our experience, providing a simple explanation of the interpretation of non-integer link weights can be challenging. In the VOSviewer software developed by two of us (Van Eck & Waltman, 2010, 2014), bibliometric networks can be constructed using either full or fractional counting, and we have the experience that users of the software sometimes have difficulties with the interpretation of non-integer link weights.

In the specific case of bibliographic coupling analysis, fractional counting may have another potential disadvantage. When using full counting, the weight of the bibliographic coupling link between two publications X and Y is fixed and cannot change over time. However, when fractional counting is used, this is no longer the case. Over time, the publications responsible for the bibliographic coupling link between publications X and Y may receive additional citations from other publications. This will then cause the weight of the bibliographic coupling link between publications X and Y to decrease. Hence, when using fractional counting, the weight of a bibliographic coupling link is no longer fixed and may decrease over time. In some situations, this might be seen as a disadvantage of fractional counting.

**4.3. Future research**

We want to emphasize that the full and fractional counting methods discussed in this paper are not the only approaches that can be taken to construct bibliometric networks. In particular, the idea of fractional counting can be extended in various ways. For instance, in a bibliographic coupling analysis of researchers, our approach is to fractionalize based on the number of citations received by a cited publication. In addition to this, one could also consider to fractionalize based on the number of citations given by a citing publication or based on the number of researchers by which a citing publication has been authored. We have not explored these possibilities in this paper, but we leave this as a topic for future research. We also note that various alternative fractionalization approaches have already been analyzed by Batagelj and Cerinšek (2013).

Future work could also study the relationship between on the one hand different approaches for constructing bibliometric networks and on the other hand different approaches for determining the similarity or relatedness of objects in these networks (e.g., Ahlgren, Jarneving, & Rousseau, 2003; Klavans & Boyack, 2006; Van Eck & Waltman, 2008, 2009). Constructing bibliometric networks and determining the similarity or relatedness of objects in these networks are two closely related problems that may benefit from being studied in a combined way.

## Acknowledgments

We would like to thank Vladimir Batagelj, Giovanni Colavizza, Vincent Traag, and three referees for valuable comments on earlier versions of this paper.

Antonio Perianes-Rodriguez worked on this research project at the Centre for Science and Technology Studies of Leiden University as awardee of a José Castillejo grant (CAS15/00178) funded by the Spanish Ministry of Education.

## References

Ahlgren, P., Jarneving, B., & Rousseau, R. (2003). Requirements for a cocitation similarity measure, with special reference to Pearson's correlation coefficient.

*Journal of the American Society for Information Science and Technology*, *54*(6), 550–560.

Aksnes, D.W., Schneider, J.W., & Gunnarsson, M. (2012). Ranking national research systems by citation indicators. A comparative analysis using whole and fractionalised counting methods. *Journal of Informetrics*, *6*(1), 36–43.

Batagelj, V. & Cerinšek, M. (2013). On bibliographic networks. *Scientometrics*, *96*(3), 845–864.

Börner, K., Chen, C., & Boyack, K.W. (2003). Visualizing knowledge domains. *Annual Review of Information Science and Technology*, *37*(1), 179–255.

Cerinšek, M. & Batagelj, V. (2015). Network analysis of Zentralblatt MATH data. *Scientometrics*, *102*(1), 977–1001.

Cronin, B. (2001). Hyperauthorship: A postmodern perversion or evidence of a structural shift in scholarly communication practices? *Journal of the American Society for Information Science and Technology*, *52*(7), 558–569.

De Solla Price, D.J. (1965). Networks of scientific papers. *Science*, *149*, 510–515.

Gauffriau, M., Larsen, P.O., Maye, I., Roulin-Perriard, A., & Von Ins, M. (2007). Publication, cooperation and productivity measures in scientific research. *Scientometrics*, *73*(2), 175–214.

Glänzel, W. & Czerwon, H.J. (1996). A new methodological approach to bibliographic coupling and its application to the national, regional and institutional level. *Scientometrics*, *37*(2), 195–221.

Kessler, M.M. (1963). Bibliographic coupling between scientific papers. *American Documentation*, *14*(1), 10–25.

Klavans, R., & Boyack, K.W. (2006). Identifying a better measure of relatedness for mapping science. *Journal of the American Society for Information Science and Technology*, 57(2), 251–263.

McCain, K.W. (1990). Mapping authors in intellectual space: A technical overview. *Journal of the American Society for Information Science*, *41*(6), 433–443.

Milojević, S. (2014). Network analysis and indicators. In Y. Ding, R. Rousseau, & D. Wolfram (Eds.), *Measuring scholarly impact: Methods and practice* (pp. 57–82). Springer.

Newman, M.E.J. (2001a). The structure of scientific collaboration networks. *Proceedings of the National Academy of Sciences of the United States of America*, *98*(2), 404–409.

Newman, M.E.J. (2001b). Scientific collaboration networks. I. Network construction and fundamental results. *Physical Review E*, *64*(1), 016131.

Newman, M.E.J. (2001c). Scientific collaboration networks. II. Shortest paths, weighted networks, and centrality. *Physical Review E*, *64*(1), 016132.

Park, H.W., Yoon, J., & Leydesdorff, L. (2016). The normalization of co-authorship networks in the bibliometric evaluation: The government stimulation programs of China and Korea. *Scientometrics*, *109*(2), 1017–1036.

Persson, O. (1994). The intellectual base and research fronts of JASIS 1986–1990. *Journal of the American Society for Information Science*, *45*(1), 31–38.

Persson, O. (2001). All author citations versus first author citations. *Scientometrics*, *50*(2), 339–344.

Persson, O. (2010). Identifying research themes with weighted direct citation links. *Journal of Informetrics*, *4*(3), 415–422.

Rousseau, R., & Zuccala, A. (2004). A classification of author co-citations: Definitions and search strategies. *Journal of the American Society for Information Science and Technology*, *55*(6), 513–529.

Sen, S.K. & Gan, S.K. (1983). A mathematical extension of the idea of bibliographic coupling and its applications. *Annals of Library Science and Documentation*, *30*(2), 78–82.

Small, H. (1973). Co-citation in the scientific literature: A new measure of the relationship between two documents. *Journal of the American Society for Information Science*, *24*(4), 265–269.

Small, H., & Sweeney, E. (1985). Clustering the Science Citation Index using co-citations: I. A comparison of methods. *Scientometrics*, *7*(3–6), 391–409.

Van Eck, N.J., & Waltman, L. (2008). Appropriate similarity measures for author cocitation analysis. *Journal of the American Society for Information Science and Technology*, *59*(10), 1653–1661.

Van Eck, N.J., & Waltman, L. (2009). How to normalize cooccurrence data? An analysis of some well-known similarity measures. *Journal of the American Society for Information Science and Technology*, *60*(8), 1635–1651.

Van Eck, N.J., & Waltman, L. (2010). Software survey: VOSviewer, a computer program for bibliometric mapping. *Scientometrics*, *84*(2), 523–538.

Van Eck, N.J., & Waltman, L. (2014). Visualizing bibliometric networks. In Y. Ding, R. Rousseau, & D. Wolfram (Eds.), *Measuring scholarly impact: Methods and practice* (pp. 285–320). Springer.

Waltman, L. (2016). A review of the literature on citation impact indicators. *Journal of Informetrics*, *10*(2), 365–391.

Waltman, L., Calero-Medina, C., Kosten, J., Noyons, E.C.M., Tijssen, R.J.W., Van Eck, N.J., ... Wouters, P. (2012). The Leiden Ranking 2011/2012: Data collection, indicators, and interpretation. *Journal of the American Society for Information Science and Technology*, *63*(12), 2419–2432.

Waltman, L., & Van Eck, N.J. (2015). Field-normalized citation impact indicators and the choice of an appropriate counting method. *Journal of Informetrics*, *9*(4), 872–894.

White, H.D., & Griffith, B.C. (1981). Author cocitation: A literature measure of intellectual structure. *Journal of the American Society for Information Science*, *32*(3), 163–171.

Zhao, D. (2006). Towards all-author co-citation analysis. *Information Processing & Management*, *42*(6), 1578–1591.

Zhao, D., & Strotmann, A. (2008a). Evolution of research activities and intellectual influences in information science 1996–2005: Introducing author bibliographic-coupling analysis. *Journal of the American Society for Information Science and Technology*, *59*(13), 2070–2086.

Zhao, D., & Strotmann, A. (2008b). Comparing all-author and first-author co-citation analyses of information science. *Journal of Informetrics*, *2*(3), 229–239.

Zhao, D., & Strotmann, A. (2011). Counting first, last, or all authors in citation analysis: A comprehensive comparison in the highly collaborative stem cell research field. *Journal of the American Society for Information Science and Technology*, *62*(4), 654–676.

Zhao, D., & Strotmann, A. (2015). *Analysis and visualization of citation networks*. Morgan & Claypool Publishers.

**Appendix**

In this appendix, we discuss how our fractional counting method for constructing co-authorship networks relates to the approaches for constructing co-authorship networks introduced by Batagelj and Cerinšek (2013) and Park et al. (2016).

In our fractional counting method, the number of co-authorship links between researchers *i* and *j* equals

$$\sum_{k=1}^{M} \frac{a_{ik} a_{jk}}{n_k - 1}, \quad (A1)$$

where *M* denotes the number of publications included in the analysis, $n_k$ denotes the number of authors of publication *k*, and $a_{ik}$ indicates whether researcher *i* is an author of publication *k* ($a_{ik} = 1$) or not ($a_{ik} = 0$).

Batagelj and Cerinšek (2013) discuss three approaches for constructing co-authorship networks. The first approach is equivalent to the conventional full counting method. The second approach is similar to our fractional counting method, but there is a minor difference. In the second approach, the number of co-authorship links between researchers *i* and *j* is given by

$$\sum_{k=1}^{M} \frac{a_{ik} a_{jk}}{n_k}. \quad (A2)$$

In Eq. (A2) the denominator equals $n_k$, while in Eq. (A1) it equals $n_k - 1$. As explained in Subsection 2.3, having a denominator of $n_k - 1$ ensures that each action has the same weight. In other words, when a researcher decides to co-author a publication, the weight of this action does not depend on the number of researchers with whom the publication is co-authored. In the case of a denominator of $n_k$, different actions may have different weights. For instance, the decision to co-author a publication with three other researchers (i.e., $n_k = 4$) introduces three co-authorship links, each with a weight of 1 / 4 = 0.25, which results in a total weight of 3 × 0.25 = 0.75. On the other hand, the decision to co-author a publication with nine other researchers (i.e., $n_k = 10$) yields nine co-authorship links, each with a weight of 1 / 10 = 0.1, resulting in a total weight of 9 × 0.1 = 0.9. In the case of a denominator of $n_k - 1$, co-authoring a publication with three other researchers results in a total weight of 3 × (1 / 3) = 1, and co-authoring a publication with nine other researchers also results in a total weight of 9 × (1 / 9) = 1. Hence, by having a denominator of $n_k - 1$, each action has exactly the same weight, while in the case of a denominator of $n_k$ different actions have the same weight only by approximation.

In the third approach of Batagelj and Cerinšek (2013), the number of co-authorship links between researchers *i* and *j* is given by

$$\sum_{k=1}^{M} \frac{a_{ik} a_{jk}}{n_k^2}. \quad (A3)$$

In this approach, the denominator equals $n_k^2$. Hence, co-authorship links calculated using Eq. (A3) have lower weights than co-authorship links calculated using Eqs. (A1) and (A2). In essence, the difference is that Eq. (A3) fractionalizes based on the total number of co-authorship links related to a publication while Eqs. (A1) and (A2) fractionalize based on the number of co-authorship links related to a single co-author of a publication. As explained above, in the case of Eqs. (A1) and (A2), each co-authorship action has (approximately) the same weight. On the other hand, using Eq. (A3), for each publication the overall weight of all co-authorship actions related to the publication is (approximately) the same, regardless of the number of authors of the publication. From the point of view of an individual researcher, this means that co-authoring a publication with a small number of other researchers has more weight than co-authoring a publication with a large number of other researchers.

Park et al. (2016) compare different approaches for constructing co-authorship networks. Their integer counting approach is equivalent to the conventional full counting method. They also study a fractional counting approach, but this approach is different from our fractional counting method. In fact, the fractional counting approach of Park et al. (2016) is identical to the third approach of Batagelj and Cerinšek (2013) discussed above.

To illustrate the differences between the various approaches for constructing co-authorship networks, we use the example introduced in Subsection 2.3. For this example, Table A1 shows the co-authorship matrix obtained using the second approach of Batagelj and Cerinšek (2013) as well as the co-authorship matrix obtained using their third approach or, equivalently, using the fractional counting approach of Park et al. (2016). For comparison, the co-authorship matrices obtained using the full counting method and our fractional counting method can be found in Table 3 in Subsection 2.3.

Table A1. Co-authorship matrices obtained using the second and the third approach of Batagelj and Cerinšek (2013).

| Second approach | R1 | R2 | R3 | R4 | Total | Third approach | R1 | R2 | R3 | R4 | Total |
|---|---|---|---|---|---|---|---|---|---|---|---|
| R1 |  | 0.33 | 0.83 | 0.00 | 1.17 | R1 |  | 0.11 | 0.36 | 0.00 | 0.47 |
| R2 | 0.33 |  | 0.33 | 0.50 | 1.17 | R2 | 0.11 |  | 0.11 | 0.25 | 0.47 |
| R3 | 0.83 | 0.33 |  | 0.00 | 1.17 | R3 | 0.36 | 0.11 |  | 0.00 | 0.47 |
| R4 | 0.00 | 0.50 | 0.00 |  | 0.50 | R4 | 0.00 | 0.25 | 0.00 |  | 0.25 |
| Total | 1.17 | 1.17 | 1.17 | 0.50 |  | Total | 0.47 | 0.47 | 0.47 | 0.25 |  |